\documentclass{optica-article}

\journal{opticajournal} 

\articletype{Research Article}

\begin{document}

\title{Future proofing network encryption technology with continuous-variable quantum key distribution}

\author{Nitin Jain,\authormark{1} Hou-Man Chin,\authormark{1} Adnan A.E. Hajomer,\authormark{1} Dev Null,\authormark{1} Henrik Larfort,\authormark{2} Naja Lautrup Nysom,\authormark{1} Erik Bidstrup,\authormark{3} Ulrik L. Andersen,\authormark{1} and Tobias Gehring\authormark{2,*}}

\address{\authormark{1}Department of Physics, Technical University of Denmark, 
2800 Kongens Lyngby, Denmark\\
\authormark{2}Energinet, Tonne Kjærsvej 65, 7000 Fredericia, Denmark\\
\authormark{3}Zybersafe ApS, Erik Husfeldts Vej 7, Denmark}

\email{\authormark{*}tobias.gehring@fysik.dtu.dk} 


\begin{abstract*} 
We demonstrate a proof-of-concept establishment of quantum-secure data transfer links in field trials at two locations in Denmark: on the campus of Technical University of Denmark in Lyngby and between power grid nodes owned and operated by Energinet in Odense. Several different links, implemented physically using optical ground wires, underground fibers as well as their combinations, were investigated. Coherent `quantum' states at 1550 nm, prepared and measured using a semi-autonomous continuous-variable quantum key distribution (CVQKD) prototype, were multiplexed in wavelength with `classical' 100Gbps encrypted data traffic from a pair of commercial layer-2 network encryption devices operating at around 1300 nm. Under the assumptions of real-time data processing, we estimate average secret key rates of $434.8, 148.6$, and $78.3$ kbps in the asymptotic limit for diverse channels with losses (at 1550 nm) of 4.1, 5.5, and 6.7 dB, respectively. 
The demonstrations permit an evaluation of the prototype's tolerance to harsh field conditions and showcase that CVQKD can serve as an additional layer to protect sensitive network traffic propagating on insecure channels.
\end{abstract*}

\section{Introduction}
The explosion in the data traffic volumes in the last few years, driven by rapid growth of data centers and cloud service providers, poses significant security challenges~\cite{cisco_annual}. This is due to the nature of the data involved, which ranges from critical national infrastructure and user identities to financial secrets such as intellectual property, all of which are at risk. In particular, the energy infrastructure has been increasingly targeted, as evidenced by incidents such as the breach at a nuclear power plant in India~\cite{india_nuc_breach}, the Colonial Pipeline ransomware case in USA~\cite{col_pipe_hack}, and most recently, the coordinated hacking of 22 different energy companies in Denmark~\cite{dk_cyberangreb}. These events highlight a worrisome trend of increasing frequency, sophistication, and involvement of nation-states in cyberattacks. 

While implementing good practices such as timely application of patches and measures to prevent exploitation of vulnerabilities can never be underscored more, cryptographic solutions such as encryption add a protective layer on the sensitive data to foil adversaries. In the industrial context, a typical approach involves the use of hardware encryption devices, also known as network encryptors or Ethernet encryptors~\cite{noauthor_ieee_2018}. 
The devices, employing symmetric cryptographic algorithms such as advanced encryption standard (AES), operate in pairs, utilizing the same sequence of pre-configured keys for encryption and decryption of data. 

One of the most secure methods to distribute random symmetric keys across an insecure communication channel is provided by quantum key distribution (QKD)~\cite{BB84in2014, Scarani2009, Pirandola2021}. In QKD, two users, Alice and Bob, can share quantum correlations through the exchange of quantum-optical signals, while ensuring that an eavesdropper, Eve, cannot intercept these correlations without getting disclosed. This is because any eavesdropping attempt introduces noise and/or loss that Alice and Bob can quantitatively estimate, assuming pre-calibrated devices. 

For QKD to become a practical cryptographic technology, one major challenge is its integration into existing optical communication infrastructure. The increased loss and noise from the telecommunication equipment, often attributed to the actions of an eavesdropper, adversely affects the key generation. Nonetheless, the QKD community has made significant progress, and several studies and field trials over the last decade have experimentally verified the feasibility of multiplexing QKD signals with classical data traffic without substantial performance degradation for either~\cite{choi_field_2014, Eriksson2019,  woodward_quantum_2021, Wang_QAN_2021, brunner_demonstration_2023}. 

In this work, we showcase an example of quantum-classical integration for securing data in transit using QKD and network encryption. A semi-autonomous, telecom-rack mountable continuous-variable (CV) QKD system~\cite{Chin2020, Jain_composable_2022, OFC2022, Chin_dig_sync_2022}, operating at 1550 nm, produced a secret keystream buffer for a pair of layer 2 network encryptors~\cite{zybersafe_cloak}, communicating simultaneously over the same channel using 100G optical transceivers at 1300 $\pm$ 10 nm. Demonstrations were conducted at two locations in Denmark with different combinations of underground and aerial fibers acting as the channel for the propagation of the C-band quantum states and the O-band optical signals, with off-the-shelf coarse wavelength division multiplexers (CWDMs) used for their union and separation. 

\begin{figure*}[t]
   \centering
    \includegraphics[width=0.97\linewidth]{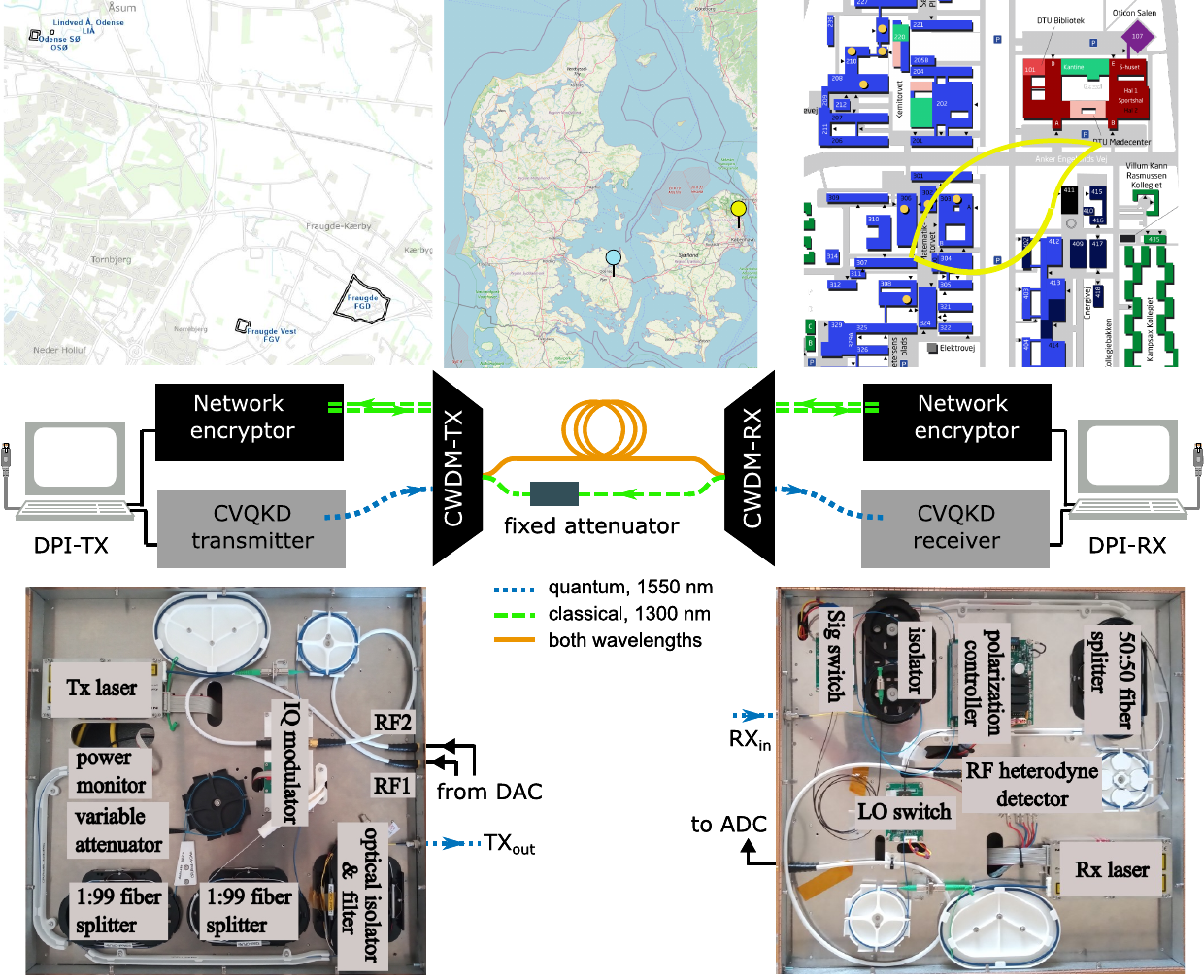}
    \caption{Practical network encryption with continuous-variable quantum key distribution. (Top row) Map of Denmark in the center, with the geographical locations for the demonstrations indicated using the cyan and yellow pins. On the left is a satellite view showing the power substation locations in Odense, and on the right is a zoomed-in map of DTU. (Center row) Optical connections are depicted using dashed-green, dotted-blue or solid-orange lines depending on the wavelength(s) of light traversing through them, with a single/pair of line(s) indicating simplex/duplex operation. A fixed attenuator and a pair of patchcords provide a path for signals at $\sim$ 1300 nm to traverse from RX to TX. (Bottom row) The CVQKD-TX and RX modules comprise a solid plate with the optical/optoelectronic components on one side (as shown) and the necessary control electronics on the other. The main text contains a detailed description of all relevant devices as well as the system operation. TX: transmitter, RX: receiver, DPI: digital processing interface.} 
    \label{fig:scheme}
\end{figure*}

Table~\ref{tab:summary} summarizes the lengths and losses (at 1550 nm) of the deployed channels, labelled L1-L4. 
\begin{table}[!t]
\footnotesize
\caption{Information about the links investigated in this work and the secret key rate (SKR) obtained thereon. All links were loopbacks in the sense that both the input and output fiber were physically accessible at the same place. The loss and length at Lyngby were estimated using simple optical power and pulse delay measurements. At Odense, these quantities were determined using an optical time domain reflectometer.} 
\label{tab:summary}
\begin{center}
\begin{tabular}{p{0.6cm}|p{1.0cm}|p{2.4cm}|p{0.9cm}|p{1.0cm}|p{1.0cm}}
    \hline
    \textbf{Link label} & \textbf{Location} & \textbf{Channel (fiber) type} & \textbf{Length, km} & \textbf{Physical loss, dB} & \textbf{SKR, kbps}\\  
    \hline
    \textbf{L1} & Lyngby & Underground only & $0.55$ & $5.5$ & $148.6$ \\
    \hline
    \textbf{L2} & Odense & Aerial only & $13.2$ & $4.1$ & $434.8$ \\ 
    \hline
    \textbf{L3} & Odense & Underground + Aerial & $15.3$ & $6.7$ & $78.3$ \\    
    \hline
    \textbf{L4} & Odense & Aerial only & $25.9$ & $8.9$ & $0.0$ \\    
\end{tabular}
\end{center}
\end{table}
The entire quantum-secure communication scheme, illustrated in 
Figure~\ref{fig:scheme}, was first implemented on L1, a loopback single-mode fiber (SMF) under the Lyngby campus of Technical University of Denmark (DTU). 
This was followed by temporary installation and operation of the QKD system at the premises of Energinet in Odense. Here, we explored three different loopback topologies L2-L4 using optical ground wires (OPGWs) running between power substations located at Fraugde and South East Odense (Odense S\O), as indicated in the top-left map in Fig.~\ref{fig:scheme}. More specifically, two of the links (L2 and L4) were a 2$\times$ and 4$\times$ back-and-forth loop of these OPGWs / aerial fibers, while L3 was a combination of an OPGW and an underground SMF. 

The last column in Table~\ref{tab:summary} shows the final results---secret key rate (SKR) estimates in the asymptotic limit---obtained on these links, assuming real-time signal and data processing at the operating rate ($B=20\,$MBaud) of our CVQKD system. In practice, keys were generated offline. The large wavelength separation (1550 vs 1300 nm) of the quantum and classical signals together with the usage of balanced coherent detection ensured that despite being more than 7 orders of magnitude higher in launch power, the classical signal did not have any observable influence on the quantum signal noise characteristics. The system's performance was challenged to some extent by the harsh environmental conditions (especially for L2-L4) which made it harder to control the channel parameters. For instance, on L4, the high untrusted loss as well as relatively high untrusted/excess noise prevented achieving a positive SKR. Nevertheless, on L1-L3, where keys could be generated successfully, the SKR values are sufficiently large to support the refresh rate requirement of 30 AES 256-bit secret key pairs per hour for continuous and real-time operation of the network encryptors. Our work thus serves as a proof-of-concept demonstration of using QKD to `future proof' network encryption technology. We note that QKD has been previously used in a deployed electric utility fiber network for secure authentication of smart grid communications~\cite{alshowkan_authentication_2022}. 

The paper is organized as follows: In the next section, we provide details of the complete quantum-classical setup that enables future proofing network encryption with CVQKD. This is followed by a description of the conducted experiments and the obtained results. 
Thereafter, we discuss the results, in particular, focusing on the channel fluctuations and their influence on the live optimization of the state of polarization (SOP) of the 1550 nm signal. Finally, we provide an outlook before concluding the paper. 
\section{Scheme: Signals, Channels \& Devices}
At the heart of the scheme shown in Fig.~\ref{fig:scheme} is the CVQKD system comprising a transmitter (TX) and receiver (RX), implemented as standard 3U 19-inch modules suitable for telecom rack installation. Coined \texttt{qTReX}~\cite{OFC2022}, this system uses randomness from a state-of-the-art quantum random number generator (QRNG)~\cite{Gehring2021} and a host of digital signal processing (DSP) methods~\cite{Kleis2017, Chin2020, Chin_dig_sync_2022} to enable the sharing of quantum correlations. Classical information processing steps, applied offline on sequences of data frames from the CVQKD-TX and -RX, utilize these correlations to distill secret keys for the encryptors. 

Currently, the entire setup can be operated with minimal user intervention. Once the classical data processing stack is also brought online, the entire operation from quantum states' production to live data encryption would become fully autonomous. With a few hardware modifications, the \texttt{qTReX} system can be adopted in data-center-like environments~\cite{qkdata_report}. 

In this section, we present the building blocks of the setup. More specifically, we elaborate on the different communication channels, the exchanged signals, the connected devices, and their roles in the different stages of the QKD protocol. For an introduction to the principles of CVQKD as well as typical CVQKD schemes, we refer the reader to reviews in Refs.~\cite{Scarani2009, Diamanti2015, Laudenbach2018, Pirandola2019}. 


\subsection{Signals}\label{scheme:sig}
The \texttt{qTReX} system uses the so-called sideband coding approach~\cite{Lance2005} for the preparation and measurement of the quantum states: Information containing complex symbols, with both the real and imaginary parts being a Gaussian random variable, was encoded into and decoded from coherent states, respectively~\cite{Chin2020, Jain_composable_2022}. 

However, the quantum-optical signal coded at the sideband frequencies is relatively dim. To address the tasks of carrier recovery and clock recovery---necessary for phase-reference alignment and synchronization of the RX and TX---the CVQKD system utilizes some ancillary signals with higher signal-to-noise ratio (SNR). These ancillary signals and the quantum-optical signal (multiplexed in time and frequency; as explained later) were the only signal connecting the CVQKD-TX and RX.  To emphasize, neither any kind of electronic triggering signal nor any reference-clock-sharing between the RX and TX was implemented.

\subsection{Quantum channel}
These signals were transported from the TX to RX via the optical connection formed by the dotted blue lines, the solid orange line, and the CWDMs in Fig.~\ref{fig:scheme}. This was the quantum channel and signals labelled TX$_{\text{out}}$ and RX$_{\text{in}}$ (shown near the bottom of the figure) further visualize its entry and exit points. We further note that the solid orange line represents the physical links L1--L4 mentioned in Table~\ref{tab:summary}. 

\subsection{Classical channel \label{scheme:chnlclass}}
While the CVQKD devices were connected solely through the quantum channel, their respective digital processing interfaces (DPIs; described in subsection~\ref{scheme:dpis}) were connected `classically'. Specifically, a pair of workstations on the local network formed a classical channel (not shown in Fig.~\ref{fig:scheme}) for the exchange of data between TX and RX during the classical phase of the QKD protocol. Ideally, this classical channel should be authenticated~\cite{Scarani2009}, typically achieved using a part of the secret key from an earlier round of the protocol. However, to simplify the overall scheme, we did not implement such authentication. 

\subsection{Encryption channel}
For high-speed (1-100 Gbps) network encryption over dedicated point-to-point links serving data centers, cloud service providers, and large enterprises, fiber-optic small form factor pluggables (SFPs) are the standard solution for distances spanning several kilometers~\cite{cisco_sfp}. 

We used off-the-shelf 100G QSFP28 4WDM-40 hot-pluggable modules~\cite{qsfp100G} for transmitting and receiving the encrypted data. As depicted by the dashed-green line pairs in Fig.~\ref{fig:scheme}, these transceiver modules operate in duplex mode. To ensure bidirectional traffic at 1300 $\pm$ 10 nm without saturating the SFP receiver at the encryptor on the TX side, we inserted a fixed attenuator (loss $\sim$ 5 dB at 1300 nm) between the CWDMs.  
The encryption channel thus constituted the quantum channel and fiber patch cords with the fixed attenuator (TX to RX: solid-orange line, RX to TX: dashed-green line). 

\subsection{CVQKD-TX}\label{schm:tx}
The output of a 1550\,nm continuous-wave laser (Tx laser) was fed to an in-phase and quadrature (IQ) modulator, biased for performing carrier suppression and single sideband modulation~\cite{Jain2021}, with the radio frequency (RF) input through ports RF1 and RF2. The RF waveforms were generated using an external digital-to-analog converter (DAC). 

\begin{figure*}[!t]
   \centering
    \includegraphics[width=0.95\linewidth]{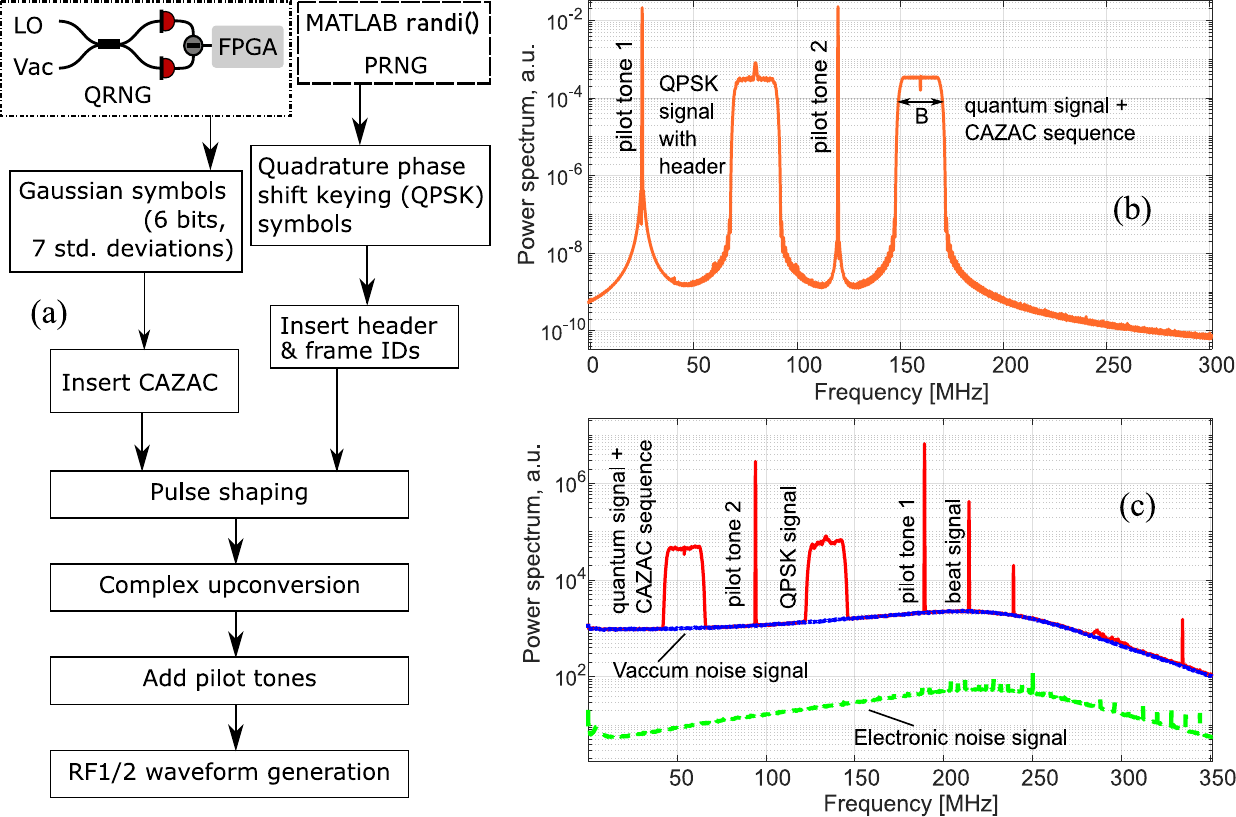}
    \caption{Digital signal processing (DSP) in \texttt{qTReX}. (a) Flowchart showing the DSP procedure for quantum state encoding in a $B=20\,$MHz wide sideband and multiplexing of  ancillary signals in time and frequency. (b) Spectrum of the complex signal, the real and imaginary parts of which are played out using a two-channel DAC. Due to time multiplexing, the quantum signal and CAZAC sequence may share the same frequency band, however, the former is $\sim 50$ times lower in power than the latter. (c) Spectra of the ADC captured signals under various configurations of the optical switches in CVQKD-RX, detailed further in the text. Vac: vacuum state, Q/PRNG: quantum/pseudo random number generator. CAZAC: constant amplitude zero auto correlation.} 
    \label{fig:dspNspectra}
\end{figure*}

The flowchart in Fig.~\ref{fig:dspNspectra}(a) illustrates the DSP steps for creating the digital input to the DAC, while the spectrum of the complex waveform, whose real (imaginary) part is related to RF1 (RF2), is depicted in Fig.~\ref{fig:dspNspectra}(b). 
Uniformly distributed random numbers from the QRNG were transformed into discrete Gaussian-distributed numbers with 6 bits of resolution and a range of 7 standard deviations~\cite{Gehring2021}. These formed the amplitudes and phases of the quantum states, and to these `quantum' symbols, a constant amplitude zero auto correlation sequence was multiplexed in time, while a quadrature phase shift keying (QPSK) symbol sequence containing a known header and frame identification (ID) information was multiplexed in frequency~\cite{Chin_dig_sync_2022}. The subsequent step of pulse shaping was performed by upsampling to the DAC sampling rate of 1 GSamples/s and applying root raise cosine filtering~\cite{Chin2020, Jain_composable_2022}. After digital single sideband modulation or complex upconversion of the two 20 MHz wide sidebands at 80 and 160 MHz, two pilot tones at 25 MHz and 120 MHz were also multiplexed in frequency as depicted in Fig.~\ref{fig:dspNspectra}(b). 

Referring to Fig.~\ref{fig:scheme} (bottom), the optical power at the output of the IQ modulator was monitored using a 1:99 fiber splitter and power meter. We could independently set the 1) modulation strength associated with the quantum states, 2) power of the pilot tones, and 3) power in the QPSK signal band to desired levels through a combined digital control of signal amplitudes during the DSP procedure, tuning of the voltage output levels of the DAC, and the optical attenuation using an electronically controlled variable optical attenuator after the IQ modulator. 

All fiber-optic components featured polarization maintaining fiber (PMF) pigtails, except the SMF-connectorized wavelength filter and isolator (added to prevent Trojan-horse attacks~\cite{Jain2014}). The signal $\text{TX}_{\text{out}}$ was fed by a simplex cable to the input of CWDM-TX.
\subsection{CVQKD-RX}\label{schm:rx}
The RX laser (same model as the TX laser, but detuned in frequency by around 215 MHz) supplied the local local oscillator (LLO) for coherent heterodyning in a balanced detector configuration. An external analog-to-digital converter (ADC) sampled the analog output of the detector at 1 GSamples/s. Remotely controlled optical (Sig and LO) switches enabled the acquisition of the IQ modulated signal coming from the quantum channel as well as calibration of the vacuum/shot noise normalization. The high extinction of over 60 dB in these switches ensured that the detector response with a switch in the OFF state is akin to physically disconnecting that fiber. 

The red trace in Fig.~\ref{fig:dspNspectra}(c) illustrates the modulated signal spectrum measured at the input of the quantum channel. It features the same multiplexed components presented in Fig.~\ref{fig:dspNspectra}(b). The spurious features on the right hand side of the beat signal were due to the finite sideband suppression~\cite{Jain2021}. With the signal beam switch in the OFF state, the ADC captured two signals featuring a variance proportional to vacuum noise plus electronic noise power (with LO switch ON) and only electronic noise power (LO switch OFF). The dotted-blue and dashed-green traces in Fig.~\ref{fig:dspNspectra}(c) show the corresponding spectra. Subtracting the two variances yields the normalization factor corresponding to the fundamental vacuum noise. 

Due to arbitrary polarization fluctuations in the channel, the incoming signal $\text{RX}_{\text{in}}$ required dynamic polarization control; section~\ref{exp:polcol} describes the details of this procedure. The signal polarization is in a desired state when it aligns with the polarization of the LLO, where after, the ADC captured the modulated signal. 

The remotely controlled switches and the dynamic polarization controller facilitated frequent and independent vacuum noise calibration and maximising the SNR of the quantum signal, respectively, thus playing a crucial role in the fast and autonomous quantum state measurement. 

\subsection{Digital Processing Interfaces}\label{scheme:dpis}
Two workstations equipped with high-end graphic processing unit (GPU) cards formed the DPIs for operating the entire QKD protocol. PCI Express based DAC and ADC cards, housed in these workstations, were used for driving the modulator in CVQKD-TX and acquiring the detector output in CVQKD-RX, respectively, as explained above. The DSP cycles running on each of them allowed the quantum symbols encoded at CVQKD-TX to be eventually decoded at CVQKD-RX, i.e., fulfil the state preparation and measurement tasks. Figure 3 in Ref.~\cite{Chin_dig_sync_2022} shows a flowchart of the DSP running on CVQKD-RX, complementary to the one Fig.~\ref{fig:dspNspectra}(a). 

Thereafter, the workstations communicated over the classical channel to perform the remaining data processing stages of the protocol, namely information reconciliation (IR), parameter estimation (PE), and privacy amplification (PA). Usage of GPUs speeds up the execution of both IR and PA~\cite{wang_IR_2018, wang_PA_2018}. 
\subsection{Network encryptors} \label{schm:nwenc}
Commercially available network encryptors from Zybersafe~\cite{zybersafe_cloak} that support the 100G Ethernet line interface and implement AES in the 256-Galois counter mode were used. The encryptors support a LAN and WAN interface for transport of the unencrypted and encrypted data, respectively. The keys are refreshed every 2 minutes or after a maximum of $2^{32}$ messages, with message size being 864 bits. 

The secret key material generated at the end of the QKD protocol was offloaded to a file, and a simple key management server running on each workstation loaded the content of these files into the encryptor using a serial connection. A pair of AES 256-bit keys (512 bits in total) at a time were supplied to the respective Ethernet encryptor for both encryption and decryption of the messages. Key IDs for these pairs were sequentially generated and exchanged on the encryption channel to ascertain that both encryptors used the correct key. 
\section{Implementation \& Results}
The goal of any QKD protocol implementation is to maximize the SKR under the given channel conditions. Before delving into the general details of the protocol runs, we first describe our efforts to overcome the challenge of polarization optimization, which plays a crucial role in the long-term and autonomous operation of \texttt{qTReX} as well as in maximisation of the SKR.  
\subsection{Dynamic polarization optimization}\label{exp:polcol}
We optimized the polarization of the incoming optical signal using the so-called simultaneous perturbation stochastic approximation (SPSA) method~\cite{spall_spsa_1992}. We used a commercial polarization controller integrated with four piezoelectric actuators with an off-the-shelf 16-bit multi-channel DAC supplying the input voltages to the drivers of the actuators. 

Controlling the SOP of the modulated signal so that it matches with the SOP of the LLO yields a stronger RF heterodyne output. The objective function to maximize was the overall power $R$ of the RF heterodyne signal, the spectrum of which is shown in Fig.~\ref{fig:dspNspectra}(c) by the red trace (with the background noise contribution shown by the blue trace). We remark here that even after channel attenuation, the total SNR was reasonably large, simply due to the various multiplexed ancillary signals. And finally, also note that the SOP of the LLO was fixed by virtue of using only PMF components, namely the Rx laser, LO switch, and 50:50 fiber splitter; see Fig.~\ref{fig:scheme}. 

The optimization was executed in a feedback loop, aiming to increase $R$ and settle around a maximum after some $N$ iterations; typically $N \sim 30$ sufficed for our system. In the supplementary section, we provide further details of how the SPSA method was exactly implemented. 
\subsection{Protocol runs: Quantum layer}\label{exp:protQ}
State machines running in a loop independently on both CVQKD-TX and CVQKD-RX used the quantum channel for transmission and reception, respectively, of the information carrying signal at 1550 nm, spectral representations of which are shown in  Fig.~\ref{fig:dspNspectra}(b) and (by the red trace) in Fig.~\ref{fig:dspNspectra}(c). Simultaneously, the network encryptors described in subsection~\ref{schm:nwenc} were set to operate continuously, using secret keys generated from prior runs of the CVQKD system, implying that the quantum channel was also inundated with the 1300 nm signal. 

At the RX, the task of polarization optimization, already described in subsection~\ref{exp:polcol} and the supplementary section, was followed by the ADC capturing samples of the modulated signal, using a level trigger set adaptively on the ADC channel itself that acquires the data. For this, the TX played out sequentially in every round a total of 1000 blocks, each containing 20 data frames, constructed using the DSP chain shown in Fig.~\ref{fig:dspNspectra}(a) with $20 \times 10^4$ unique Gaussian symbols utilized per block. This burst from the TX lasted just around 3 mins, after which, it was left in an idle state until the start of the next round, which had a duration of 10 minutes. 

While the RX was programmed to capture and store---as a file on disk---a maximum of 100 of these data blocks, in practice, the yield was typically less due to timeouts (set-trigger-level too high) or due to capturing of noise between successive TX blocks (set-trigger-level too low). After acquiring the data with modulation, a total of $2 \times 2 =$ 4G ADC samples for performing the task of vacuum noise calibration, described in section~\ref{schm:rx}, were captured. The switches routing the Sig and LO away from the detector for this purpose were then reverted to their original positions, and CVQKD-RX was then also left in an idle state until the next round. The duty cycle at the RX was typically between $70-80\%$. 

Once the data files from all 6 rounds were available, DSP routines based on a machine learning framework~\cite{Chin2020} operating at DPI-RX kicked in. Using the ancillary signals, the digital synchronization procedure was performed. In particular, the pilot tones assisted in clock and carrier recovery, while the QPSK signal helped in optimal sampling~\cite{Chin2020, Chin_dig_sync_2022} by making use of a sequence of symbols known to both TX and RX. After extracting the frame IDs, the demodulation of the QPSK band was completed. With all the frequency, phase, and temporal information obtained, the Gaussian quantum symbols could also be optimally extracted. 

\begin{figure*}[t]
   \centering
\includegraphics[width=0.92\linewidth]{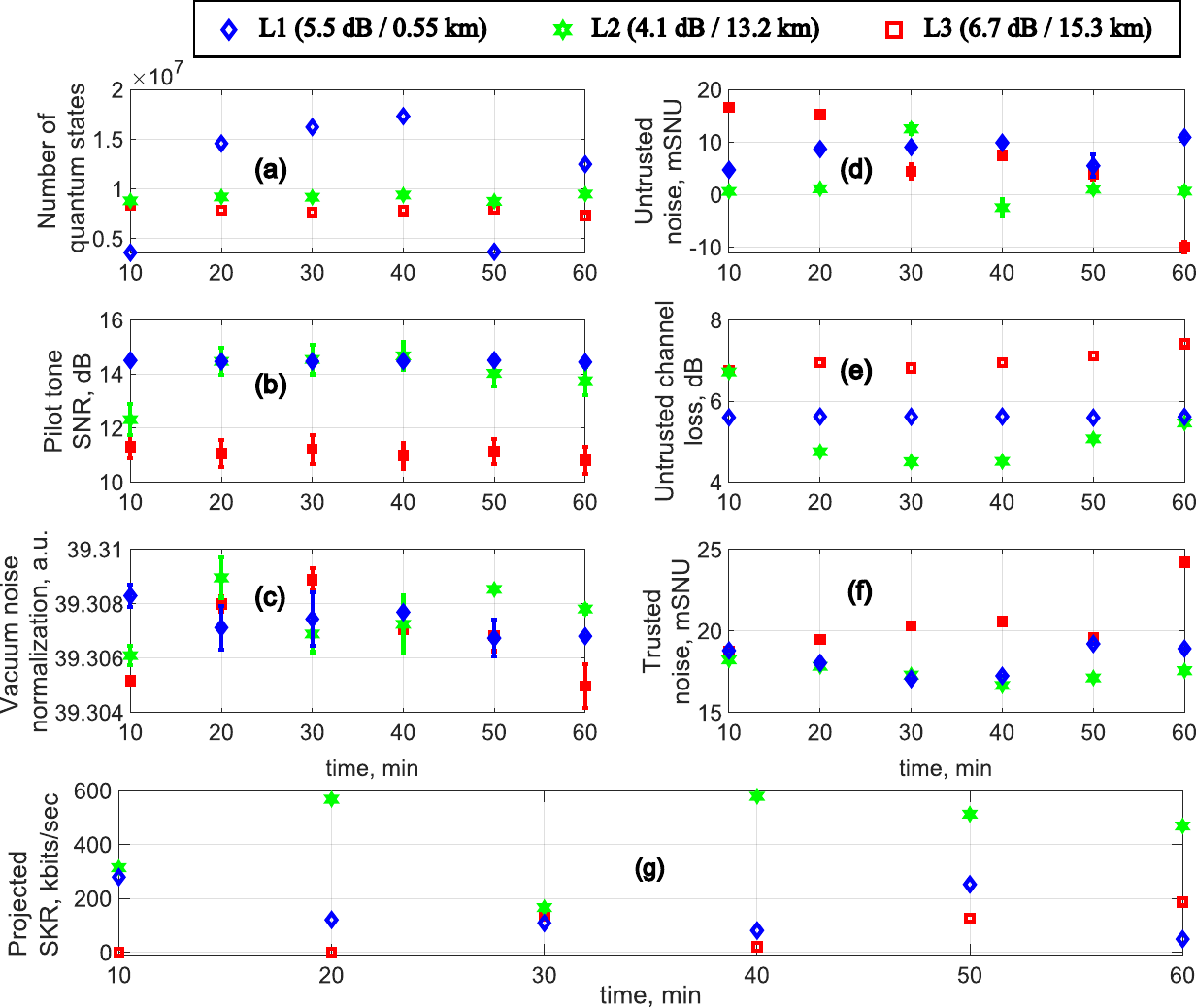}
    \caption{(Color online) Results from the parameter estimation and calibration procedures performed in the tests carried out at links L1, L2 and L3, where a positive SKR was eventually obtained (refer Table~\ref{tab:summary} for more info). (a) Number of quantum states successfully measured by RX. (b) Strength of the pilot tone which aids in the clock and carrier recovery. (c) Outcome of the calibration procedure performed after every round of modulation and acquisition. (d) and (e) Noise and loss attributed to Eve. (f) Noise assumed to be beyond Eve's control. (g) Projected secret key rate (SKR) if the DSP and information reconciliation had been performed in real time. A reconciliation efficiency $\beta = 95\%$ and FER = 0.5 was uniformly assumed in the calculations, and for L1 and L2, it was actually also achieved (further details in the main text). The PE was actually performed using all the symbols of Alice and Bob.}
    \label{fig:compiled_res}
\end{figure*}

Figure~\ref{fig:compiled_res} shows various results from experiments conducted with L1--L3 forming a dominant part of the quantum channel\footnote{The reasons for excluding results from L4 are explained towards the end of this section.}. 
Out of the $2 \times 10^8$ quantum states generated at the TX every round, Fig.~\ref{fig:compiled_res}(a) shows the number of quantum states that were successfully measured at the RX. Figures~\ref{fig:compiled_res}(b) and (c) show respectively the evolution of the measured SNR of the pilot tone and the main outcome of the vacuum noise calibration procedure described above. At the TX, while the quantum phase of the protocol was $6 \times 10 = 60$ mins long, at the RX, it took around 3 times longer as the DSP was not performed in real time. 
\subsection{Protocol runs: Classical layer}\label{exp:protC}
With the quantum stage of the protocol finished, the classical data processing steps of IR, PE, and PA were also performed offline. The IR was based on a multi-dimensional reverse reconciliation scheme and a bank of multi-edge-type low-density-parity-check codes were used~\cite{mani_phd_2020}. For IR, Alice filtered her entire symbol train based on the captured frame ID information, shared by Bob via the classical channel. Due to a limited number of code rates, high-performance IR was actually possible only on data from L1 and L2. In these cases, we successfully obtained a reconciliation efficiency $\beta \approx 95\%$ at a frame error rate (FER) in the vicinity of 0.5.  

Alice then performed PE using her filtered symbols and the quantum symbols of Bob recovered by her through IR decoding, together with the variables such as the electronic and vacuum noise variances from the calibrations in RX. After calculating the covariance matrix, the total noise $\xi$ and transmittance $\eta$ were evaluated. These could be partitioned into trusted and untrusted components~\cite{Jain_composable_2022} using the vacuum noise normalization values shown in Fig.~\ref{fig:compiled_res}(c), and the optical transmittance of the receiver (including CWDM-RX) $\eta_t = 0.33$. Figures~\ref{fig:compiled_res}(d) and (e) show the untrusted components $\eta_u = \eta/\eta_t$ of the transmittance (converted to loss in dB) and noise $\xi_u = \xi - \xi_t$ respectively, from the 6 rounds, while the trusted noise values $\xi_t$ themselves are shown in Fig.~\ref{fig:compiled_res}(f). Note that these noise figures are as measured, i.e., referred to the output. As is often done in CVQKD, the noise is conveyed as whatever is in excess of the shot/vacuum noise in a relative sense, i.e., in shot noise unit (SNU). Together, these quantities aid in the calculation of the SKR, given by 
\begin{equation}\label{eq:skr}
\text{SKR} = B (1 - \text{FER}) \left( \beta I_{AB} - \chi_{BE} \right),  
\end{equation}
where the mutual information $I_{AB}$ between Alice and Bob is also calculated using the covariance matrix. The level of Eve's correlations to Bob's data is typified by $\chi_{BE}$, the Holevo information. From a security perspective, the keys generated based on this formula are secure in the asymptotic regime, and assume collective attacks by Eve~\cite{Grosshans2003, Diamanti2015}. 

Figure~\ref{fig:compiled_res}(g) shows the projected SKR values for L1-L3, which vary mainly due to the fluctuations of the channel parameters $\eta_u$ and $\xi_u$ during the 6 rounds, and in some cases, also result in the SKR crossing the null threshold ($\beta I_{AB} = \chi_{BE}$). In fact, in case of L4, the values of $\eta_u$ and $\xi_u$ forced $\beta I_{AB} < \chi_{BE}$ for all rounds, due to which the results from L4 are not plotted in Figure~\ref{fig:compiled_res}. Finally, also note that due to statistical uncertainties $\xi_u < 0$ in Fig.~\ref{fig:compiled_res}(d) in a couple of cases; for the SKR evaluation, we replaced this with a slightly positive value of 2 mSNU. 
\section{Discussion} \label{sec:disc}
Collecting statistics over the 6 rounds depicted in Fig.~\ref{fig:compiled_res}(g) yields  the SKR values as $148.6 \pm 94.3$, $434.8 \pm 162.8$, and $78.3 \pm 81.1$ kbps for L1, L2, and L3, respectively. In this section, we discuss the reasons behind these large variations in the SKR as well as how they can be brought under control. Furthermore, we also provide justification for the assumptions that the DSP, IR, and PA is performed in real time and that it is always possible to obtain $\beta \gtrsim 95\%$ and FER $\lesssim$ 0.5 in IR. With these assumptions, such rates can indeed be achieved in practice. 

We also remark that a stronger notion of composability~\cite{Jain_composable_2022} of the secret key could not be satisfied here due to the insufficient number of quantum states obtained in the experiments here as well as the relatively worse trusted loss ($\eta_t = 0.33$ for \texttt{qTReX} versus $\eta_t = 0.68$ in Ref.~\cite{Jain_composable_2022}). 
\subsection{Channel loss: Untrusted vs Physical}\label{disc:loss}
The channel loss values graphed in Fig.~\ref{fig:compiled_res}(e) over the 6 rounds show more fluctuation for links L2 and L3 than L1. Furthermore, comparing with the physical loss values presented in Table~\ref{tab:summary}, a difference of at least $0.3\,$dB is evident. 
The latter can partly be explained due to the fixed loss coming from two extra fiber adapters that were used for connecting to the loopbacks at Odense; since they were absent during the OTDR measurements, their combined loss is not included in the entries of Table~\ref{tab:summary}. 

The remaining deviation as well as the fluctuation stems primarily from the subpar performance of the SOP optimization. This is discussed in further detail in subsection~\ref{disc:SNRvar} below. From the perspective of channel loss estimation, a poor SOP optimization reduces the efficiency of mode-matching the LLO with the optical signal (containing both the quantum signal and the ancillary signals; see subsection~\ref{schm:rx}). This can make the execution of the overall DSP chain~\cite{Chin_dig_sync_2022} at RX inefficient as well: Below a certain pilot SNR, a cascading decrease in performance through the chain is obtained, which worsens the carrier and clock recovery, causing the PE evaluated noise and/or loss to rise. 
\subsubsection{SNR variations}\label{disc:SNRvar} 
\begin{figure*}[t]
   \centering
\includegraphics[width=0.95\linewidth]{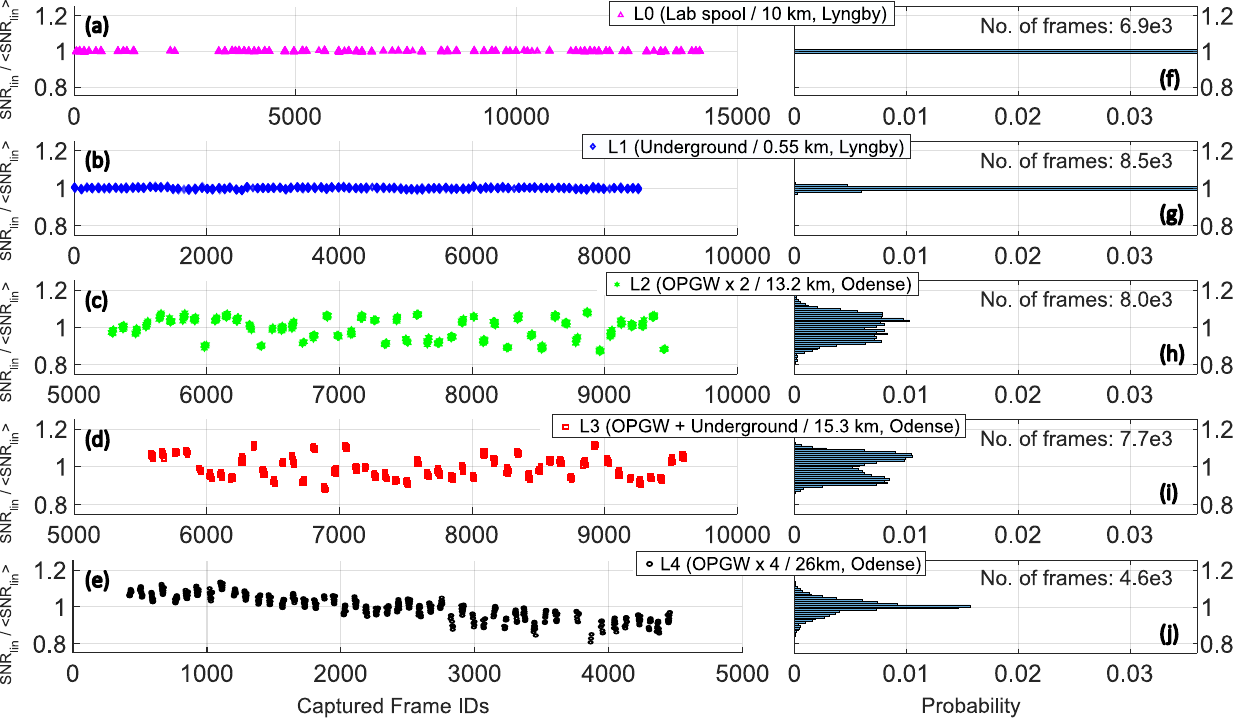}
    \caption{(Color online) Fluctuations of the relative SNR estimated from the QPSK signal for quantum channels L0-L4. The relative SNR, denoted by $\text{SNR}_{\text{lin}}$ / $\langle \text{SNR}_{\text{lin}} \rangle$, is calculated as the ratio of linearized SNR values from a ($\leq 10~$mins long) round to the mean value computed over that entire round, and allows an unbiased comparison among all the channels, where the operating SNRs were otherwise widely different. (a)-(e) Relative SNRs as a function of the IDs of the successfully captured frames from one exemplary round. Each data point in these 5 subplots reflects the SNR calculated from a total of $20$K QPSK symbols, constituting one frame. The frame ID sequence differs in each subplot because of the usage of adaptive trigger level, explained in Section~\ref{exp:protQ}, which causes the successful-trigger-instants to differ in every round (and thus, obviously also across different channels). (f)-(j) Respective histograms drawn using the relative SNRs from all 6 rounds; the distributions thus reflect the relative SNR variations over a period of 1 hour. The bin width is uniformly chosen to be 0.01 for all links.} 
    \label{fig:snr_fluct}
\end{figure*}

Figure~\ref{fig:snr_fluct} gives a deeper insight into the varying characteristics of the links (L1-L4) investigated in this work by comparing them with similar results from a 10 km long lab fiber spool (L0), investigated before in Ref.~\cite{OFC2022}. 
More specifically, Figs.~\ref{fig:snr_fluct}(a)-\ref{fig:snr_fluct}(e) show the variation of the QPSK signal SNR from a single round of the QKD protocol, executed right after the dynamic SOP optimization procedure explained in Section~\ref{exp:polcol}. The corresponding normalized histograms in Figs.~\ref{fig:snr_fluct}(f)-\ref{fig:snr_fluct}(j) for the channels are plotted using data accumulated from all 6 rounds, with the number of elements for the histograms, i.e., frames, mentioned in the respective legends. 

As outlined in Section~\ref{exp:protQ}, a block of 20 frames is periodically modulated by the transmitter: Due to this, chunks with up to 20 frames/block may be seen in Figs.~\ref{fig:snr_fluct}(a)-\ref{fig:snr_fluct}(e). An inter-block variation of the SNR, which occurs on a few $100$s of millisecond scales, is much more pronounced for L2-L4 than either L0 or L1. A vertical elongation of these chunks, indicating much faster fluctuations (on a $1-10\,$ms scale), is most prominent for the longest channel L4. These could be explained by the fact that the rate of change of the SOP of a lightwave in OPGWs can be much higher than in underground fibers, as the former are exposed to not only larger and faster temperature variations but also the natural forces of wind and lightning~\cite{Charlton2017}. 

Any intra and inter-block variations, together with variations from one round to another, result in the distributions for L2, L3 and L4 being significantly wider than for L0 and L1, as illustrated in Figs.~\ref{fig:snr_fluct}(f)-\ref{fig:snr_fluct}(j). We conjecture that the different physical locations (Odense vs Lyngby; see Table~\ref{tab:summary}) of the QKD devices play a role in the broadening of the distributions. To elaborate, the QKD devices, specifically the electronic interface of the polarization controller inside CVQKD-TX (see subsection~\ref{exp:polcol}), were subjected to a high degree of electromagnetic interference (EMI) emanating from the power transformers and grid-lines in case of L2-L4\footnote{We note here that measurement results (not shown in Fig.~\ref{fig:snr_fluct}) from a B2B configuration at Odense also indicate a wider distribution, however the collected data was an order of magnitude lower, which hinders a reliable statistical conclusion.}. In contrast, the experiments conducted over L0 and L1 had the devices placed in a lab environment with a significantly lower EMI. 

This implies that the SPSA method used for optimizing polarization works fairly well for links such as L0 and L1, but to actively compensate for fluctuations in L2-L4, the SOP optimization should be run more frequently. Also, with a longer and more thorough characterization of the aerial fibers, a better choice of the initial optimization parameters may help; however, this was not possible due to time constraints. In a future work, it would also be interesting to accurately identify the individual contributions of the SOP fluctuations and high-noise environment to these SNR variations. Finally, an alternative that does not require the implementation of any active method for SOP optimization is to use a so-called polarization hybrid~\cite{Pereira2023, Chin_polhyb_2023}, where the fluctuations can be taken into account during DSP. 
\subsection{Practical and real-time data processing}\label{disc:pracproc}
As presented in the previous section, the bottlenecks in a real time implementation of the CVQKD protocol stack primarily come from the DSP at RX as well as the IR and PA procedures. Also, for a fully autonomous operation, the data processing pipelines at both TX and RX require synchronization. 

We are currently working on some major revisions in our DSP software chain and in the near future, also plan to port the DSP routines on an FPGA where we expect a considerable speedup. Furthermore, we already have an IR framework capable of performing multidimensional reconciliation as well as privacy amplification in near real time. To elaborate, on an Intel Core i7-9750H CPU (with 6 cores and 2.60 GHz as the processor base frequency) and an NVIDIA GeForce RTX 2060 Mobile GPU (with 1920 cores and 6 GB memory), the IR and PA throughput can reach 9.85 MBaud, which is almost half of the raw symbol rate. All in all, we believe with all these advances, it will be possible to achieve the data processing pipeline for the CVQKD protocol to operate close to $B=20\,$MBaud in the near future. 

The main challenge that then remains to solve is how to pick the appropriate input parameters such as the code rate, the log likelihood ratio factor, etc. that ensure a single-shot reconciliation with the simultaneous assurance of a high $\beta$ at a reasonable FER. The overall IR performance is fairly sensitive to the choice of these parameters, which also need to be frequently adjusted in case of channels with a fluctuating transmittance~\cite{mani_phd_2020}. For this, we are currently working on an algorithm that uses the ancillary signals to enable a dynamic calculation of these input parameters which can then make the IR part of the classical data processing stack meet all the aforementioned requirements. This would additionally facilitate a feedback between TX and RX to optimize the modulation strength, thus maximising the SKR for the given channel parameters. 
\section{Conclusion \& Outlook}
In conclusion, we have reported field trials of a semi-autonomous continuous variable quantum key distribution (CVQKD) system that powers a network encryption scheme to establish quantum-secure data transfer links at two locations in Denmark. More specifically, we successfully operated the CVQKD system outside the lab, with the devices located in a high electromagnetic noise environment of a power substation and performing quantum communication over combinations of underground and aerial fibers as the quantum channels. The CVQKD system operated at 1550 nm while the network encryptors operated at 1300 nm and used the same (wavelength-multiplexed) channel for the transport of both quantum states as well as encrypted data. Despite the several orders of magnitude difference in optical powers, the scheme was able to have a positive secret key yield on 3 of the 4 investigated channels, exhibiting another successful quantum-classical integration example. This showcases the technological readiness level of QKD for protecting critical infrastructure data in a reliable and robust way. Furthermore, such a quantum-classical platform can be used to future proof the security of network data infrastructure, such as data center interconnections. 
\section*{Acknowledgements}
We would like to thank Danish e-infrastructure Cooperation (DeiC) for the equipment loan as well as the use of the loopback fiber at DTU. We are also thankful to Morten Houborg Andersen, Alexander Frederiksen, and Thomas Wisbech from Energinet for valuable discussions as well as facilitating the equipment installation at the substation in Fraugde. We acknowledge support from Innovation Fund Denmark (CryptQ project, grant agreement no.\ 0175-00018A) and the Danish National Research Foundation, Center for Macroscopic Quantum States (bigQ, DNRF142). This project has received funding from the European Union’s Digital Europe programme under Grant agreement No 101091659 (QCI.DK).
\section*{Disclosures}
The authors declare no conflicts of interest. 
\section*{Data availability} 
Data underlying the results presented in this paper are not publicly available at this time but may be obtained from the authors upon reasonable request. 

\bibliography{references}

\begin{thebibliography}{10}
\newcommand{\enquote}[1]{``#1''}

\bibitem{cisco_annual}
\enquote{{Cisco} {Annual} {Internet} {Report} (2018–2023) {White} {Paper},}
  Weblink:
  \url{https://www.cisco.com/c/en/us/solutions/collateral/executive-perspectives/annual-internet-report/white-paper-c11-741490.html}.

\bibitem{india_nuc_breach}
J.~M. Porup, \enquote{{How a nuclear plant got hacked},}
  \url{www.csoonline.com/article/568145/how-a-nuclear-plant-got-hacked.html}.
  Accessed: 2023-11-23.

\bibitem{col_pipe_hack}
\enquote{{The Colonial Pipeline Hack Is a New Extreme for Ransomware},}
  \url{https://www.wired.com/story/colonial-pipeline-ransomware-attack/}.
  Accessed: 2023-11-23.

\bibitem{dk_cyberangreb}
\enquote{{The attack against Danish, critical infrastructure},}
  \url{https://sektorcert.dk/wp-content/uploads/2023/11/SektorCERT-The-attack-against-Danish-critical-infrastructure-TLP-CLEAR.pdf}.
  Accessed: 2023-11-25.

\bibitem{noauthor_ieee_2018}
\enquote{{IEEE} {Standard} for {Local} and metropolitan area networks-{Media}
  {Access} {Control} ({MAC}) {Security},} {\protect\JournalTitle{IEEE Std
  802.1AE-2018 (Revision of IEEE Std 802.1AE-2006)}} pp. 1--239 (2018).

\bibitem{BB84in2014}
C.~H. Bennett and G.~Brassard, \enquote{Quantum cryptography: Public key
  distribution and coin tossing,} {\protect\JournalTitle{Theoretical Computer
  Science}} \textbf{560}, 7--11 (2014). Theoretical Aspects of Quantum
  Cryptography – celebrating 30 years of BB84.

\bibitem{Scarani2009}
V.~Scarani, H.~Bechmann-Pasquinucci, N.~J. Cerf, \emph{et~al.}, \enquote{{The
  security of practical quantum key distribution},}
  {\protect\JournalTitle{Reviews of Modern Physics}} \textbf{81}, 1301--1350
  (2009).

\bibitem{Pirandola2021}
S.~Pirandola, \enquote{{Limits and security of free-space quantum
  communications},} {\protect\JournalTitle{Physical Review Research}}
  \textbf{3}, 013279 (2021).

\bibitem{choi_field_2014}
I.~Choi, Y.~R. Zhou, J.~F. Dynes, \emph{et~al.}, \enquote{Field trial of a
  quantum secured 10 {Gb}/s {DWDM} transmission system over a single installed
  fiber,} {\protect\JournalTitle{Optics Express}} \textbf{22}, 23121--23128
  (2014). Publisher: Optica Publishing Group.

\bibitem{Eriksson2019}
T.~A. Eriksson, T.~Hirano, B.~J. Puttnam, \emph{et~al.}, \enquote{{Wavelength
  division multiplexing of continuous variable quantum key distribution and
  18.3 Tbit/s data channels},} {\protect\JournalTitle{Communications Physics}}
  \textbf{2}, 9 (2019).

\bibitem{woodward_quantum_2021}
R.~I. Woodward, J.~F. Dynes, P.~Wright, \emph{et~al.}, \enquote{Quantum {Key}
  {Secured} {Communications} {Field} {Trial} for {Industry} 4.0,} in \emph{2021
  {Optical} {Fiber} {Communications} {Conference} and {Exhibition} ({OFC}),}
  (2021), pp. 1--3.

\bibitem{Wang_QAN_2021}
B.-X. Wang, S.-B. Tang, Y.~Mao, \emph{et~al.}, \enquote{Practical quantum
  access network over a 10 gbit/s ethernet passive optical network,}
  {\protect\JournalTitle{Opt. Express}} \textbf{29}, 38582--38590 (2021).

\bibitem{brunner_demonstration_2023}
H.~H. Brunner, C.-H.~F. Fung, M.~Peev, \emph{et~al.}, \enquote{Demonstration of
  a switched {CV}-{QKD} network,} {\protect\JournalTitle{EPJ Quantum
  Technology}} \textbf{10}, 1--12 (2023). Number: 1 Publisher: SpringerOpen.

\bibitem{Chin2020}
H.-M. Chin, N.~Jain, D.~Zibar, \emph{et~al.}, \enquote{{Machine learning aided
  carrier recovery in continuous-variable quantum key distribution},}
  {\protect\JournalTitle{npj Quantum Information}} \textbf{7}, 20 (2021).

\bibitem{Jain_composable_2022}
N.~Jain, H.-M. Chin, H.~Mani, \emph{et~al.}, \enquote{Practical
  continuous-variable quantum key distribution with composable security,}
  {\protect\JournalTitle{Nature Communications}} \textbf{13}, 4740 (2022).

\bibitem{OFC2022}
N.~Jain, H.-M. Chin, H.~Mani, \emph{et~al.}, \enquote{{qTReX : A
  semi-autonomous continuous-variable quantum key distribution system},} More
  information at \url{ https://ieeexplore.ieee.org/document/9748181}.

\bibitem{Chin_dig_sync_2022}
H.-M. Chin, N.~Jain, U.~L. Andersen, \emph{et~al.}, \enquote{Digital
  synchronization for continuous-variable quantum key distribution,}
  {\protect\JournalTitle{Quantum Science and Technology}} \textbf{7}, 045006
  (2022).

\bibitem{zybersafe_cloak}
\enquote{{Zybersafe TrafficCloak -- Datasheet},}
  \url{https://zybersafe.com/wordpress/wp-content/uploads/2019/11/Zybersafe-Data-Sheet.pdf}.
  Accessed: 2023-01-03.

\bibitem{alshowkan_authentication_2022}
M.~Alshowkan, P.~G. Evans, M.~Starke, \emph{et~al.}, \enquote{Authentication of
  smart grid communications using quantum key distribution,}
  {\protect\JournalTitle{Scientific Reports}} \textbf{12}, 12731 (2022).

\bibitem{Gehring2021}
T.~Gehring, C.~Lupo, A.~Kordts, \emph{et~al.}, \enquote{{Homodyne-based quantum
  random number generator at 2.9 Gbps secure against quantum
  side-information},} {\protect\JournalTitle{Nature Communications}}
  \textbf{12}, 1--11 (2021).

\bibitem{Kleis2017}
S.~Kleis, M.~Rueckmann, and C.~G. Schaeffer, \enquote{{Continuous variable
  quantum key distribution with a real local oscillator using simultaneous
  pilot signals},} {\protect\JournalTitle{Optics Letters}} \textbf{42},
  1588--1591 (2017).

\bibitem{qkdata_report}
N.~Jain, T.~Gehring, U.~Hoff, \emph{et~al.}, \enquote{{Using QKD for Data
  Center Security},}
  \url{https://www.copenhagenfintech.dk/projects/using-qkd-for-data-center-security}.
  Accessed: 2023-12-10.

\bibitem{Diamanti2015}
E.~Diamanti and A.~Leverrier, \enquote{{Distributing secret keys with quantum
  continuous variables: Principle, security and implementations},}
  {\protect\JournalTitle{Entropy}} \textbf{17}, 6072--6092 (2015).

\bibitem{Laudenbach2018}
F.~Laudenbach, C.~Pacher, C.-H.~F. Fung, \emph{et~al.},
  \enquote{{Continuous-Variable Quantum Key Distribution with Gaussian
  Modulation-The Theory of Practical Implementations},}
  {\protect\JournalTitle{Advanced Quantum Technologies}} \textbf{1}, 1800011
  (2018).

\bibitem{Pirandola2019}
S.~Pirandola, U.~L. Andersen, L.~Banchi, \emph{et~al.}, \enquote{{Advances in
  quantum cryptography},} {\protect\JournalTitle{Advances in Optics and
  Photonics}} \textbf{12}, 1012 (2020).

\bibitem{Lance2005}
A.~M. Lance, T.~Symul, V.~Sharma, \emph{et~al.}, \enquote{{No-Switching Quantum
  Key Distribution Using Broadband Modulated Coherent Light},}
  {\protect\JournalTitle{Physical Review Letters}} \textbf{95}, 180503 (2005).

\bibitem{cisco_sfp}
\enquote{{A move to high speed server {connectivity} in the cloud},}
  \url{https://www.cisco.com/c/en/us/products/collateral/interfaces-modules/transceiver-modules/high-speed-server-con-cloud-wp.html}.
  Accessed: 2023-11-03.

\bibitem{qsfp100G}
\enquote{{QSFP28 100G 4WDM-40 transceiver},}
  \url{https://apps.juniper.net/hct/model/?component=QSFP-100G-4WDM40}.
  Accessed: 2023-11-03.

\bibitem{Jain2021}
N.~Jain, I.~Derkach, H.~M. Chin, \emph{et~al.}, \enquote{{Modulation leakage
  vulnerability in continuous-variable quantum key distribution},}
  {\protect\JournalTitle{Quantum Science and Technology}} \textbf{6} (2021).

\bibitem{Jain2014}
N.~Jain, B.~Stiller, I.~Khan, \emph{et~al.}, \enquote{{Risk analysis of Trojan
  - horse attacks on practical quantum key distribution systems},}
  {\protect\JournalTitle{IEEE Journal on Selected Topic in Quatum Electronics}}
  \textbf{21}, 1077--260X (2014).

\bibitem{wang_IR_2018}
X.~Wang, Y.~Zhang, S.~Yu, and H.~Guo, \enquote{High speed error correction for
  continuous-variable quantum key distribution with multi-edge type {LDPC}
  code,} {\protect\JournalTitle{Scientific Reports}} \textbf{8}, 10543 (2018).

\bibitem{wang_PA_2018}
X.~Wang, Y.~Zhang, S.~Yu, and H.~Guo, \enquote{High-{Speed} {Implementation} of
  {Length}-{Compatible} {Privacy} {Amplification} in {Continuous}-{Variable}
  {Quantum} {Key} {Distribution},} {\protect\JournalTitle{IEEE Photonics
  Journal}} \textbf{10}, 1--9 (2018).

\bibitem{spall_spsa_1992}
J.~Spall, \enquote{Multivariate stochastic approximation using a simultaneous
  perturbation gradient approximation,} {\protect\JournalTitle{IEEE
  Transactions on Automatic Control}} \textbf{37}, 332--341 (1992).

\bibitem{mani_phd_2020}
H.~Mani, \enquote{Error {Reconciliation} {Protocols} for
  {Continuous}-{Variable} {Quantum} {Key} {Distribution},} Ph.D. thesis,
  Technical University of Denmark (2020).

\bibitem{Grosshans2003}
F.~Grosshans, G.~{Van Assche}, J.~Wenger, \emph{et~al.}, \enquote{{Quantum key
  distribution using gaussian-modulated coherent states},}
  {\protect\JournalTitle{Nature}} \textbf{421}, 238--241 (2003).

\bibitem{Charlton2017}
D.~Charlton, S.~Clarke, D.~Doucet, \emph{et~al.}, \enquote{Field measurements
  of sop transients in opgw, with time and location correlation to lightning
  strikes,} {\protect\JournalTitle{Opt. Express}} \textbf{25}, 9689--9696
  (2017).

\bibitem{Pereira2023}
D.~Pereira, A.~N. Pinto, and N.~A. Silva, \enquote{Polarization diverse true
  heterodyne receiver architecture for continuous variable quantum key
  distribution,} {\protect\JournalTitle{Journal of Lightwave Technology}}
  \textbf{41}, 432--439 (2023).

\bibitem{Chin_polhyb_2023}
H.-M. Chin, A.~A. Hajomer, N.~Jain, \emph{et~al.}, \enquote{Machine learning
  based joint polarization and phase compensation for cv-qkd,} in \emph{2023
  Optical Fiber Communications Conference and Exhibition (OFC),}  (2023), pp.
  1--3.

\end{thebibliography}

\end{document}